# The Fast Heuristic Algorithms and Post-Processing Techniques to Design Large and Low-Cost Communication Networks

Yahui Sun [iD], Marcus Brazil [iD], Doreen Thomas [iD], *Senior Member, IEEE*, and Saman Halgamuge [iD], *Fellow, IEEE*

*Abstract*—It is challenging to design large and low-cost communication networks. In this paper, we formulate this challenge as the prize-collecting Steiner Tree Problem (PCSTP). The objective is to minimize the costs of transmission routes and the disconnected monetary or informational profits. Initially, we note that the PCSTP is MAX SNP-hard. Then, we propose some post-processing techniques to improve suboptimal solutions to PCSTP. Based on these techniques, we propose two fast heuristic algorithms: the first one is a quasilinear time heuristic algorithm that is faster and consumes less memory than other algorithms; and the second one is an improvement of a state-of-the-art polynomial time heuristic algorithm that can find high-quality solutions at a speed that is only inferior to the first one. We demonstrate the competitiveness of our heuristic algorithms by comparing them with the state-of-the-art ones on the largest existing benchmark instances (169 800 vertices and 338 551 edges). Moreover, we generate new instances that are even larger (1 000 000 vertices and 10 000 000 edges) to further demonstrate their advantages in large networks. The state-of-the-art algorithms are too slow to find high-quality solutions for instances of this size, whereas our new heuristic algorithms can do this in around 6 to 45s on a personal computer. Ultimately, we apply our post-processing techniques to update the best-known solution for a notoriously difficult benchmark instance to show that they can improve near-optimal solutions to PCSTP. In conclusion, we demonstrate the usefulness of our heuristic algorithms and post-processing techniques for designing large and low-cost communication networks.

## I. INTRODUCTION

THE emerging next-generation communication networks, including the 5G wireless networks [1], the cognitive radio networks [2] and the Internet of Things [3], are envisioned to have a great impact on our future society, economy,

Manuscript received July 1, 2017; revised March 10, 2018 and July 5, 2018; accepted December 11, 2018; approved by IEEE/ACM TRANSACTIONS ON NETWORKING Editor Y. Zhang. This work was supported by the Melbourne International Research Scholarship of the University of Melbourne and the Australian Research Council under Grant LP140100670. *(Corresponding author: Yahui Sun.)*

Y. Sun is with the Research School of Engineering, Australian National University, Canberra, ACT 2601, Australia (e-mail: yahui.sun@anu.edu.au).

M. Brazil and D. Thomas are with the School of Electrical, Mechanical and Infrastructure Engineering, The University of Melbourne, Parkville, VIC 3010, Australia (e-mail: brazil@unimelb.edu.au; doreen.thomas@unimelb.edu.au).

S. Halgamuge is with the School of Electrical, Mechanical and Infrastructure Engineering, The University of Melbourne, Parkville, VIC 3010, Australia, and also with the Research School of Engineering, Australian National University, Canberra, ACT 2601, Australia (e-mail: saman.halgamuge@anu.edu.au).

This paper has supplementary downloadable material available at http://ieeexplore.ieee.org, provided by the authors.

Digital Object Identifier 10.1109/TNET.2018.2888864

and quality of life. These networks are expected to have an unprecedented large scale, and may be too expensive without carefully designing the network topology to make a good trade-off between minimizing the costs of transmission routes and maximizing the connected monetary or informational profits. Thus, making such a trade-off to design large and low-cost communication networks is a challenge that needs to be addressed.

Mathematically, the challenge of designing low-cost communication networks can be formulated as Steiner tree problems (e.g. designing the routing [4]–[6] or physical [7]–[9] topologies of low-cost communication networks). In this paper, we formulate the challenge of designing low-cost communication networks as the Prize-Collecting Steiner Tree Problem (PCSTP), where we are given a connected undirected graph with vertices and edges; each vertex is associated with a nonnegative node weight (or prize); each edge is associated with a positive cost; and the objective is to find a connected subgraph to minimize the included edge costs plus the missed node weights. By using vertices to represent communication spots (e.g. houses or sensors); using edges to represent transmission routes between spots (wired or wireless); using node weights to represent prizes that can be earned from spots (e.g. monetary profits or sensing information); and using edge costs to represent costs of transmission routes (e.g. wired cable costs or wireless energy consumption costs), we can design low-cost communication networks by solving PCSTP and therefore making a good trade-off between minimizing the costs of transmission routes and maximizing the connected monetary or informational profits. For example, in Figure 1, we can minimize the costs of transmission routes and the missed monetary or informational profits in communication networks.

In fact, the techniques for PCSTP have already been applied fruitfully by AT&T to the optimization of real-world telecommunication networks [10]. Nevertheless, the existing techniques may not have a good performance in designing the emerging larger communication networks. For example, the largest benchmark instance they have ever challenged has only 169,800 vertices and 338,551 edges [11], which may not be large enough to represent the emerging next-generation communication networks, such as those with millions of nodes enabled by the 5G wireless and Internet of Things technology [3], [12]. Hence, there is value in developing new techniques that still perform well in larger networks.

PCSTP is a more general version of the classical Steiner tree problem in graphs. Since the classical Steiner tree problem in graphs is NP-hard [13], PCSTP is also NP-hard, which means that there is no algorithm to solve it to optimality in polynomial time unless P = NP. Like the other NP-hard problems, different solution techniques have been developed for PCSTP in small and large instances. Pre-processing techniques and exact algorithms are often combined together to produce optimal solutions in small instances (e.g. the combination in [14]), where pre-processing techniques reduce instance sizes and exact algorithms produce optimal solutions in the reduced instances (note that, the state-of-the-art pre-processing techniques, such as those in [15], are only fast enough to reduce small instances with thousands of vertices, while we deal with much larger instances with millions of vertices in this paper). On the other hand, heuristic algorithms and post-processing techniques are often combined together to produce fast suboptimal solutions in large instances (e.g. the combination in [16]), where heuristic algorithms produce fast suboptimal solutions and post-processing techniques improve these solutions. Since the next-generation communication networks are envisioned to have a large scale, we focus on heuristic algorithms and post-processing techniques in this paper.

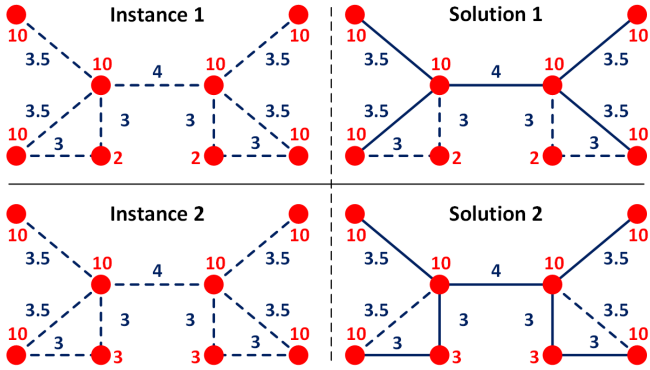

Fig. 1. Two communication network design instances. The red nodes represent communication spots; the red numbers represent prizes that can be earned from these spots; the blue dash lines represent candidate transmission routes; the blue numbers represent costs of these routes; and the blue solid lines represent transmission routes in the minimum-cost communication networks (Steiner Minimum Trees).

Many heuristic algorithms have been proposed in the last few decades. Nevertheless, the existing heuristic algorithms may still not be fast enough, and their solutions may not have a sufficiently low cost either. Moreover, post-processing techniques have so far been rarely explored for PCSTP. The Minimum Spanning Tree (MST) technique [17], which was initially proposed for the classical Steiner tree problem in graphs, is the only post-processing technique that can be used for PCSTP to date. The computational trials show that it is ineffective in many cases [16]. Thus, more powerful heuristic algorithms and post-processing techniques are both required to design large and low-cost communication networks. To address this issue, we make the following contributions:

- we propose a pruning algorithm, a growing algorithm, and then combine them with the existing MST technique together as a post-processing procedure for PCSTP.
- based on the proposed pruning algorithm, we propose the only quasilinear time heuristic algorithm for

PCSTP to date. Then, we improve a state-of-the-art polynomial time heuristic algorithm for PCSTP.

- we indicate the competitiveness of our heuristic algorithms over the state-of-the-art ones for both the largest existing benchmark instances and some newly generated instances that are even larger. Furthermore, we apply our post-processing techniques to update the best-known solution for a notoriously difficult benchmark instance to show that they can improve near-optimal solutions to PCSTP.

## II. The Current Steiner Tree Algorithms

The earliest version of PCSTP was proposed by Segev [18] in 1987. The term "Prize-Collecting Steiner Tree" was first used by Bienstock *et al.* [19] in 1993. Similar to the other NP-hard problems, the algorithms for PCSTP can be divided into two groups: exact and heuristic. Exact algorithms can find optimal solutions, but their running time does not scale well for large instances. On the other hand, heuristic algorithms cannot guarantee optimality, but they can find fast feasible solutions in large instances.

The state-of-the-art exact algorithms solve PCSTP using various Mixed Integer Programming (MIP) models. For example, Ljubić et al. [14] formulated a branch-and-cut model in 2006, and they successfully solved PCSTP to optimality in some instances with thousands of vertices. Furthermore, Fischetti *et al.* [20] proposed a node-based model. They combined the branch-and-cut model above and the node-based model together as a Steiner tree solver: Staynerd [21], which won most of the categories for PCSTP in the latest DIMACS Implementation Challenge on Steiner tree problems [22]. Recently, Leitner *et al.* [23] proposed a dual-ascent-based branch-and-bound algorithm, which performs even better than Staynerd in widely-used benchmark instances. These exact algorithms can easily find optimal or near-optimal solutions to PCSTP. However, they have high demands on machines and are slow in large instances (as are MIP-based heuristic algorithms). In particular, they consume a lot of memory and may not be able to solve large instances when there are limits on computation resources. Moreover, sophisticated MIP solvers are needed to implement these algorithms. As a result, it is not easy to embed these algorithms into mobile platforms, which may be necessary in 5G cellular networks or wireless sensor networks. Therefore, the use of simple fast heuristic algorithms is required in some cases.

The first heuristic algorithm for PCSTP was proposed by Bienstock et al. [19] in 1993, and their algorithm achieved an approximation guarantee of 3. In 1995, Goemans and Williamson [24] proposed a primal-dual algorithm using Bienstock's LP relaxation model, and their algorithm achieved an approximation guarantee of $2 - 1/(|V| - 1)$. This algorithm deals with instances where there is exactly one compulsory terminal, which is called the root. In the instances with multiple compulsory terminals, this algorithm can be implemented by selecting one compulsory terminal to be the root and giving the other ones large node weights; in the instances with no compulsory terminal, this algorithm can be implemented by trying all the possible roots. Consequently, the time complexity of this algorithm is $O(|V|^2 log|V|)$ in the instances with compulsory terminals, and $O(|V|^3 log|V|)$ in the instances with no compulsory terminal [16]. In this paper, we refer to this algorithm as the rooted GW algorithm.

There are two phases in the rooted GW algorithm, which are called GW-growth and GW-pruning. A raw solution tree is obtained in the GW-growth phase, while the GW-pruning phase improves the raw solution tree by deleting expensive edges and vertices. Johnson et al. [16] improved the rooted GW algorithm in 2000 by proposing the Strong Pruning algorithm to replace the original GW-pruning algorithm. The Strong Pruning algorithm is more effective in practice, and its known theoretical guarantees are the same [10]. Therefore, the Strong Pruning algorithm is recommended to be used to replace the original GW-pruning algorithm by many researchers [11], [16]. Moreover, they further proposed an unrooted version of the algorithm, which has a time complexity of $O(|V|^2 log|V|)$ in the instances with an arbitrary number of compulsory terminals, and its approximation guarantee is 2 [25]. The unrooted GW algorithm is much faster than the rooted one in instances with no compulsory terminal, and thus is widely used to design large networks.

Besides the early work above, Archer *et al.* [10] proposed an improved approximation algorithm in 2011 by combining the rooted GW algorithm with an MIP-based approximation algorithm for the classical Steiner tree problem in graphs [26]. The approximation guarantee of their algorithm is less than 1.9672. To our knowledge, this is the tightest approximation guarantee obtained for PCSTP so far, which is useful from a theoretical perspective. Nevertheless, it may not be fast enough in large instances, especially when considering the fact that implementing this algorithm in a rooted graph still needs to run the rooted GW algorithm twice and the MIP-based approximation algorithm once.

In this paper, we focus on designing large and low-cost communication networks. Hence, we are interested in the fast implementation of GW algorithms. The most important work to accelerate GW algorithms was that of Cole *et al.* [27] in 2001. They introduced the idea of dynamic edge splitting, which induces a time complexity of $O(|E|log^2|V|)$ for unrooted instances. Based on this idea, Hegde *et al.* [11] further accelerated the unrooted GW algorithm in 2014 by employing priority queues, and their algorithm achieves a time complexity of $O(|E|log|V|)$. We refer to this algorithm as the FGW (Fast Goemans-Williamson) algorithm, which can be considered as a state-of-the-art heuristic algorithm for PCSTP. Nonetheless, we will later show that we can improve it.

## III. THE PRIZE-COLLECTING STEINER TREE PROBLEM

In this section, we first formally define PCSTP to formulate the large and low-cost communication network design problem. Then, we indicate the difficulty to develop heuristic algorithms for PCSTP by showing that it is MAX SNP-hard.

*Definition 1 (The Prize-Collecting Steiner Tree Problem):* Let $G(V, E, C, w, c)$ be a connected undirected graph, where $V$ is the set of vertices, $E$ is the set of edges, $C$ is a (possibly empty) subset of $V$ called compulsory terminals, $w$ is a function which maps each vertex in $V$ to a nonnegative value called the node weight (or prize), and $c$ is a function which maps each edge in $E$ to a positive value called the edge cost. The purpose is to find a connected subgraph $G'(V', E'), C \subseteq V' \subseteq V, E' \subseteq E$ with the minimum net-cost $c(G') = \sum_{v \in V \setminus V'} w(v) + \sum_{e \in E'} c(e)$.

Clearly, if we let $V$ be the set of communication spots; let $E$ be the set of candidate transmission routes; let $C$ be the set of special spots that must be connected (e.g. the base stations in telecommunication or wireless sensor networks); let $w$ be the set of prizes that can be earned from spots; let $c$ be the set of costs of transmission routes, then we can design minimum-cost communication networks by solving PCSTP on $G$, and $G'$, which is the designed communication network, will have the minimum cost $c(G') = \sum_{v \in V \setminus V'} w(v) + \sum_{e \in E'} c(e)$, i.e., the sum of missed prizes and included costs is minimal.

Bern and Plassmann [28] proved that the classical Steiner tree problem in graphs with edge lengths 1 and 2 is MAX SNP-hard. Since PCSTP is a more general case of this problem, PCSTP is also MAX SNP-hard.

*Theorem 1: PCSTP is MAX SNP-hard.*

Arora *et al.* [29] showed that if any MAX SNP-hard problem has a polynomial time approximation scheme, then P = NP. Thus, there is no polynomial time approximation scheme for PCSTP unless P = NP (notably, the P versus NP problem is a major unsolved problem in computer science, and it has not yet been proven that P $\neq$ NP). Consequently, it is hard to develop fast heuristic algorithms that can produce near-optimal solutions in large instances. In this paper, we will propose two heuristic algorithms to meet this challenge: a quasilinear time one without approximation guarantee (the later proposed MSTG algorithm) and a polynomial time one with a constant approximation guarantee (the later proposed FGW′ algorithm). We will show that they are fast and can produce near-optimal solutions in large instances.

## IV. THE PROPOSED POST-PROCESSING TECHNIQUES

It may only be possible to produce suboptimal solutions to PCSTP in large instances. Here, we propose some post-processing techniques to improve suboptimal solutions to PCSTP. Our later proposed heuristic algorithms are based on these techniques.

### A. The Proposed General Pruning Algorithm

Steiner tree problems in graphs are generally NP-hard. However, some special cases of them are polynomially solvable, such as the Shortest Path Problem [30] and the Minimum Spanning Tree Problem [31]. Here, we propose the Node-Weighted Steiner Tree Problem in Trees (NWSTPT) for the first time. NWSTPT is polynomially solvable, and solving it is equivalent to improving suboptimal solutions to PCSTP by pruning expensive vertices and edges. We propose the General Pruning Algorithm (GPrA; Algorithm 1) to solve NWSTPT to optimality in polynomial time.

There are two equivalent definitions of a tree: 1) a tree is a connected network with no cycle; 2) a tree is a connected network such that the removal of any edge in this network will make it disconnected. NWSTPT is defined as follows:

**Algorithm 1** The Proposed General Pruning Algorithm (GPrA)

---

**Input:** Tree $T(V, E, C, w, c)$
**Output:** Subtree $T_p \subseteq T$
1: $T_p = T$
2: **if** $C$ is empty **then**
3:    Initialize $nw$ values
4:    Mark all the vertices as unprocessed
5:    **while** there is more than one unprocessed vertex **do**
6:      **for** unprocessed vertex $i$ such that $\xi(i) = 1$ **do**
7:        Find the unprocessed adjacent vertex $j$
8:        **if** $c(i, j) < nw(i)$ **then**
9:          Update $nw(j)$ using Equation (1)
10:        **end if**
11:        Mark vertex $i$ as processed
12:      **end for**
13:    **end while**
14:    Add the vertex with the largest $nw$ value to $C$
15: **end if**
16: Initialize $nw$ values
17: Mark all the vertices as unprocessed
18: Randomly select a compulsory terminal as the root $r$
19: **while** there is more than one unprocessed vertex **do**
20:    **for** unprocessed non-root vertex $i$ that $\xi(i) = 1$ **do**
21:      Find the unprocessed adjacent vertex $j$
22:      **if** $nw(i) < c(i, j)$ **then**
23:        Remove $(i, j)$ and the subtree rooted at $i$
24:      **else**
25:        Update $nw(j)$ using Equation (1)
26:      **end if**
27:      Mark vertex $i$ as processed
28:    **end for**
29: **end while**

---

*Definition 2 (The Node-Weighted Steiner Tree Problem in Trees):* Let $T(V, E, C, w, c)$ be a tree, where $V$ is the set of vertices, $E$ is the set of edges, $C$ is a (possibly empty) subset of $V$ called compulsory terminals, $w$ is a function which maps each vertex in $V$ to a real value called the node weight, and $c$ is a function which maps each edge in $E$ to a positive value called the edge cost. The purpose is to find a subtree $T_p(V_p, E_p), C \subseteq V_p \subseteq V$, $E_p \subseteq E$ with the maximum net-weight $w(T_p) = \sum_{v \in V_p} w(v) - \sum_{e \in E_p} c(e)$ or the minimum net-cost $c(T_p) = \sum_{v \in V \setminus V_p} w(v) + \sum_{e \in E_p} c(e)$.

When all the node weights are non-negative, NWSTPT can be considered as a special case of PCSTP where the input graph $G$ is a tree. If we consider $T$ as a suboptimal solution to PCSTP, then finding $T_p$ is to improve this solution by pruning expensive vertices and edges. The Strong Pruning algorithm proposed by Johnson *et al.* [16] in 2000 can solve NWSTPT to optimality in trees with a single compulsory terminal. We first modify this algorithm to solve NWSTPT with multiple compulsory terminals, and this modified version is incorporated into GPrA as Steps 16-29.

Suppose $C$ is nonempty, then Steps 2-15 will be skipped. We associate each vertex with an $nw$ value (Step 16). The initial $nw$ values of non-compulsory vertices are their node weights, while that of compulsory terminals are $B = \sum_{(j,k) \in E} |c(j, k)| + \sum_{j \in V} |w(j)|$, which ensures that all the compulsory terminals are included in the pruning solution. The $nw$ values will be updated in the pruning process. Note that, these $nw$ values are different from the similar values in Johnson's Strong Pruning algorithm. With these new $nw$ values, we can solve NWSTPT to optimality in trees with multiple compulsory terminals.

We further define the processing degree of vertex $i$, $\xi(i)$, as the number of its adjacent vertices that are unprocessed. Initially, all the vertices are unprocessed (Step 17), and only leaves have a processing degree of 1. We randomly select a compulsory terminal to be the root $r$ (Step 18). For unprocessed non-root vertex $i$ such that $\xi(i) = 1$, we find its unprocessed adjacent vertex $j$ (Step 21). If $nw(i) < c(i, j)$, then we remove edge $(i, j)$ and the subtree rooted at $i$ (Step 23), otherwise we update the $nw$ value of $j$ using the following equation (Step 25).

$$nw(j) = nw(j) + nw(i) - c(i, j) \qquad (1)$$

We keep processing all the non-root vertices until all of them have been processed (Step 19). This process is an improvement of Johnson's Strong Pruning algorithm since it can deal with instances with multiple compulsory terminals. However, it is not enough to solve NWSTPT or post-process PCSTP since there may be no compulsory terminal at the first stage (e.g. the instances in [14]).

Then, suppose $C$ is empty. We go through Steps 2-15 to add a non-compulsory vertex to $C$: we first initialize the $nw$ values (Step 3) in the same way above, and keep processing all the vertices that have a processing degree of 1 (Steps 6-12); this process ends when there is only one vertex left unprocessed (Step 5); the vertex associated with the largest $nw$ value is added to $C$ (Step 14). Then, we go through Steps 16-29 to prune the tree. The theorem below is proposed to prove that GPrA can solve NWSTPT to optimality in trees with an arbitrary number of compulsory terminals.

*Theorem 2:* Let $T$ be a tree with an arbitrary number of compulsory terminals, and let $T'$ be any subtree of $T$ that contains all the compulsory terminals. If $T_p$ is the subtree obtained from $T$ by GPrA, then $T_p$ contains all the compulsory terminals, and $w(T_p) \geq w(T')$.

*Proof:* We first prove this theorem for instances with compulsory terminals. Assume a compulsory terminal $r$ has been selected to be the root. Since the root will never be processed, the compulsory terminal $r$ will not be removed from $T$. Assume that there is a subtree $T_i$ in $T$ which contains at least one compulsory terminal other than $r$ and such that vertex $i$ is the last vertex being processed in this subtree. Let vertex $j$ be the predecessor of $i$ ($j$ is not in this subtree). We call $T_i$ a successor of $j$. It can be seen from Equation (1) that, after processing all the other vertices in $T_i$ (excluding $i$), $nw(i) > c(i, j)$. Thus, this subtree will be kept in $T$. As a result, $T_p$ contains all the compulsory terminals. We define a tree that is a successor of a given vertex $j$ as a Concrete Tree if it is maximal under the condition that no vertex can be removed from it by GPrA, i.e., there is no other successor of $j$ satisfying this condition while contains this tree as a subtree. Assume that there is a subtree in $T$ that has a root at vertex $q$, and there are $x$ Concrete Trees in this subtree that are successors to $q$ (see Figure 2). Vertex $q$ is connected to vertex $m_j$ in Concrete Tree $CT_j$. After processing all the vertices in these Concrete Trees, let $nw(m_\alpha) \geq c(q, m_\alpha)$, $\forall \alpha = 1, \cdots, j$, and $nw(m_\alpha) < c(q, m_\alpha), \forall \alpha = j+1, \cdots, x$, where $m_\alpha$ is the vertex in Concrete Tree $CT_\alpha$. It is easy to see that

$$w(q \cup \sum\nolimits_{\alpha=1}^{j}[(q, m_\alpha) \cup CT_\alpha])$$
$$\geq w(q \cup \sum\nolimits_{\alpha=1}^{x} z_\alpha[(q, m_\alpha) \cup CT_\alpha]|_{z_\alpha=0,1}) \quad (2)$$

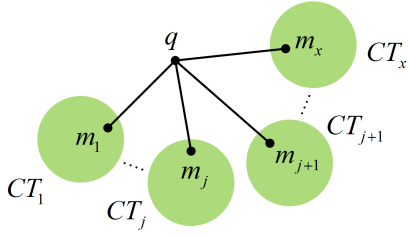

Fig. 2. A subtree rooted at vertex $q$. $CT_\alpha|_{\alpha=1,\cdots,x}$ are $x$ Concrete Trees that are successors to vertex $q$. Vertex $q$ is connected to vertex $m_j$ in Concrete Tree $CT_j$.

The Concrete Trees $CT_\alpha|_{\alpha=j+1,\cdots,x}$ will be removed from $T$ by GPrA, and the remaining part, $q \cup \sum_{\alpha=1}^{j}[(q, m_\alpha) \cup CT_\alpha]$, is also a Concrete Tree. Hence, after processing all the non-root vertices in $T$, the successor of any vertex will be a Concrete Tree. If vertex $q$ is the root $r$, then $q \cup \sum_{\alpha=1}^{j}[(q, m_\alpha) \cup CT_\alpha]$ is $T_p$. It can be seen from Equation (2) that $w(T_p) \geq w(T')$.

We then prove this theorem for instances with no compulsory terminal. Assume that $T_{opt}$ is a subtree of $T$, and $w(T_{opt}) \geq w(T')$. Therefore, adding any vertex to $T_{opt}$ or removing any vertex from $T_{opt}$ will decrease its net-weight. After Steps 3-13, there will be one vertex left unprocessed. There are two possible cases: Case 1: the unprocessed vertex is not in $T_{opt}$. Assume that vertex $i$ is the last processed vertex in $T_{opt}$. Then, $T_{opt}$ will be obtained as the subtree rooted at vertex $i$ in the process of Steps 3-13 (suppose that expensive vertices and edges are deleted in the same way as Steps 16-29), and $nw(i) = w(T_{opt})$. Case 2: the unprocessed vertex is in $T_{opt}$. Assume that vertex $i$ is that vertex. Consider $i$ as the root of $T$. $T_{opt}$ will be the remaining tree after Steps 3-13 (suppose again that expensive vertices and edges are deleted in the same way as Steps 16-29), and $nw(i) = w(T_{opt})$. Therefore, in any of these two cases, there is a vertex $i$ in $T_{opt}$ for which $nw(i) = w(T_{opt})$, and $nw(i) \geq w(T')$. It is easy to see that, after Steps 3-13, the net-weight of a Concrete Tree rooted at any vertex $j$ must be larger than or equal to $nw(j)$. If there is a vertex $j$ for which $nw(j) > nw(i)$, then the net-weight of the Concrete Tree rooted at $j$ must be larger than $w(T_{opt})$, which conflicts with the assumption. Thus, $nw(i) \geq nw(j), \forall j \in V$. By setting the vertex with the largest $nw$ value as the root of $T$ (Step 14), $T_p = T_{opt}$. $\qquad\square$

Clearly, the time complexity of GPrA is $O(|V|)$. Johnson's Strong Pruning algorithm has the same time complexity for trees with compulsory terminals but a larger time complexity

---

**Algorithm 2** The Proposed Tree Growing Algorithm (TGA)

**Input:** Graph $G(V, E, C, w, c)$, tree $T \subseteq G$, parameter $n$
**Output:** Tree $T_g \subseteq G$
1: Initialize $T_g = T$
2: Mark all the vertices in $T$ as unchecked
3: **while** there is at least one unchecked vertex **do**
4:    **for** unchecked vertex $i$ **do**
5:       **if** there is a path candidate rooted at $i$ that has a non-negative value and a length not larger than $n$ **then**
6:          Add this path to $T_g$
7:          Mark the newly added vertices as unchecked
8:       **else**
9:          Mark $i$ as checked
10:       **end if**
11:    **end for**
12: **end while**

---

of $O(|V|^2)$ for trees with no compulsory terminal (by trying all the possible roots). Since there is no compulsory terminal in most instances for PCSTP (e.g. the instances in [11] and [14]), it is preferable to apply GPrA to prune suboptimal solutions to PCSTP.

### B. The Proposed Tree Growing Algorithm

GPrA improves suboptimal solutions to PCSTP by pruning expensive vertices and edges. Here, we propose the Tree Growing Algorithm (TGA; Algorithm 2) to improve suboptimal solutions to PCSTP by adding profitable branches.

A path is a tree that only contains two leaves. Given the initial graph $G$ and a suboptimal solution tree $T$, we define a path candidate as a path in $G$ that contains only one vertex in $T$. The length of a path candidate is the number of vertices it contains that are not in $T$. The value of a path candidate is the net-weight of this path minus the node weight of the vertex that is in $T$. Clearly, path candidates with positive values are profitable branches that can be added to improve suboptimal solutions to PCSTP.

If $G$ has $|V|$ vertices and $T$ has $|S|$ vertices, the maximum number of possible path candidates is

$$P_{max} = |S| \times [\frac{(|V| - |S|)!}{0!} + \frac{(|V| - |S|)!}{1!} + \frac{(|V| - |S|)!}{2!} \cdots$$
$$+ \frac{(|V| - |S|)!}{(|V| - |S| - 2)!} + \frac{(|V| - |S|)!}{(|V| - |S| - 1)!}] \quad (3)$$

In the worst case, we need to check every path candidate to find one with a positive value. However, it is computationally too expensive to do so in large graphs as the number of possible path candidates may be quite large. In TGA, a parameter $n$ is used as the upper bound of the length of possible path candidates that are checked, and $n \geq 1$. We consistently find and add all the possible path candidates that have a non-negative value and a length not larger than $n$. The time complexity of TGA is $O(|V|^{n+1})$. It is easy to see that TGA with a larger $n$ has a higher probability of improving a suboptimal solution, but at the cost of a longer running time. We will later investigate this trade-off through computational trials.

---

**Algorithm 3** The Proposed Post-Processing Procedure (P3)

---

**Input:** Graph $G(V, E, C, w, c)$, tree $T \subseteq G$, parameter $n$
**Output:** Tree $T' \subseteq G$
1: Initialize $T' = T$
2: **while** $T'$ can be improved **do**
3:      $T' = \text{TGA}(T', G, n)$
4:      $T' = \text{MST}(T', G)$
5:      $T' = \text{GPrA}(T')$
6: **end while**

---

### C. The Proposed Post-Processing Procedure

It is hard to obtain optimal solutions to PCSTP in large instances. Therefore, post-processing techniques may be required to improve suboptimal solutions. The MST technique is the only technique to date that can post-process suboptimal solutions to PCSTP. It does this by finding the MST that spans all the vertices in that solution. Here, we combine the MST technique with the proposed GPrA and TGA together as the Post-Processing Procedure (P3; Algorithm 3). P3 can improve not only suboptimal solutions of our heuristic algorithms, but also those of the state-of-the-art exact algorithms.

Given the initial graph $G$ and a suboptimal solution tree $T$, we iteratively improve $T$ until $T$ cannot be improved any more: use TGA to improve $T$; find the MST of $T$; use GPrA to prune $T$. Note that, in the iteration above, the MST technique can improve the result of TGA, but it cannot improve the pruning result of GPrA, which is a subtree of an MST. The following theorem is well-known and proves this.

*Theorem 3:* Any subtree of an MST is an MST that spans all the vertices in that subtree.

*Proof:* Let $T$ be an MST in graph $G$, $T_s$ be a subtree of $T$, and $T_{\overline{s}}$ be the part of $T$ excluding $T_s$. Let $T_m$ be an MST in graph $G$ that spans all the vertices in $T_s$. If $c(T_m) < c(T_s)$, then $c(T_{\overline{s}} \cup T_m) < c(T)$, which conflicts with the assumption that $T$ is an MST in graph $G$. Therefore, $T_s$ is an MST in graph $G$ that spans all the vertices in it. □

The MST can be found using Prim's algorithm [32], which has a time complexity of $O(|E| + |V|log|V|)$. Therefore, the time complexity of P3 is $O(|V|^{n+1})$, where $n \geq 1$ is the parameter in TGA.

We define the domain of $T$ as a domain that contains all the subtrees of $T$. A solution tree is considered as the subtree of itself. It is easy to see that a solution tree may belong to different domains, and a domain may belong to another domain. Using GPrA to prune $T$ is to find the best solution tree in the domain of $T$. Using TGA to grow $T$ is to find a better solution tree in a domain that contains the domain of $T$. While finding the MST of $T$ is to find a better solution tree in a different domain which may or may not share common trees with the domain of $T$. The process of P3 is illustrated in Figure 3.

Clearly, P3 can be applied to instances with an arbitrary number of compulsory terminals. Note that, since the classical Steiner tree problem in graphs is a special case of PCSTP, P3 can also be used to post-process suboptimal solutions to it. However, since all the node weights are zero in the

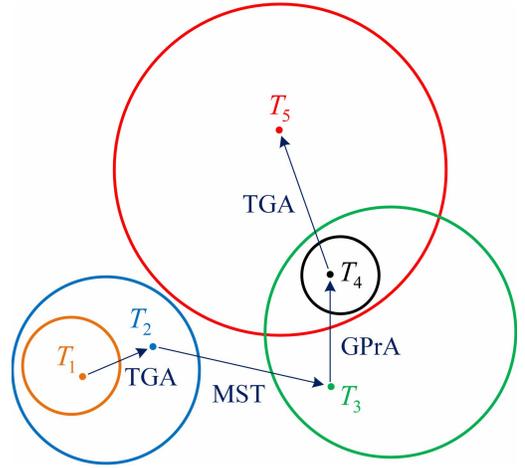

Fig. 3. The illustration of the process of P3. Each circle represents a domain. Each dot represents a solution tree. P3 improves the solution from $T_1$ to $T_5$.

---

**Algorithm 4** The Proposed MSTG Algorithm

---

**Input:** Graph $G(V, E, C, w, c)$
**Output:** Steiner tree $T \subseteq G$
1: $T = \text{MST}(G)$
2: $T = \text{GPrA}(T)$

---

classical Steiner tree problem in graphs, TGA can then be removed from P3. In this case, the time complexity of P3 is $O(|E| + |V|log|V|)$. Moreover, it is worth mentioning that P3 can even be applied to instances with negative node weights, which widely exist in some applications (e.g. [33]).

### V. A Fast Heuristic Algorithm Based on the Proposed Post-Processing Techniques

It is challenging to design large and low-cost communication networks since solving PCSTP in large instances may take a long time and consume a lot of memory. Therefore, it is preferable to develop heuristic algorithms that are fast and consume little memory to meet this challenge. In this section, we propose a quasilinear time heuristic algorithm without approximation guarantee (Algorithm 4). It is faster and consumes less memory than other algorithms.

In this algorithm, we first find the MST of the input graph, then we use GPrA to prune that MST to obtain a feasible solution. We refer to this algorithm as the MSTG algorithm (MST + GPrA). The smallest time complexity to find MST is $O(|E| + |V|log|V|)$ [32], while the time complexity of GPrA is $O(|V|)$. Therefore, the MSTG algorithm has a quasilinear time complexity of $O(|E| + |V|log|V|)$. To our knowledge, it is the only quasilinear time heuristic algorithm for PCSTP to date. There is no approximation guarantee for it (the rigorous proof is provided in the supplement). Nevertheless, it may still be preferable to apply the MSTG algorithm in some cases since

- it has advantages over the other algorithms in time or computational resource sensitive scenarios as it is faster and consumes less memory.

- it can solve extremely large instances that the other algorithms are too slow to solve.
- it can find reasonably high-quality solutions in practice (we will later show this through computational trials).

## VI. The Proposed Fast Implementation of Goemans-Williamson Algorithms

Our MSTG algorithm above is an extremely fast heuristic algorithm without approximation guarantee. However, in some cases, it may be preferable to apply slower heuristic algorithms with approximation guarantee to design large and low-cost communication networks. The improved GW algorithms are the state-of-the-art polynomial time heuristic algorithms with approximation guarantee. Their improvements can be divided into two groups: one is to improve the approximation ratio, the other one is to accelerate the optimization process. The improvements with better approximation ratios help us to understand PCSTP in a different way. Nevertheless, these algorithms may not be fast enough in large instances. Moreover, these algorithms may not be able to find high-quality solutions in practice, even though they can guarantee the absence of low-quality solutions. In this section, we focus on the latter group of improvements. We propose an improved fast implementation of the unrooted GW algorithm, and a fast implementation of the rooted GW algorithm. Ultimately, we analyze the solution certainty of GW approximation scheme for the first time.

### A. The Improved Fast Implementation of the Unrooted Goemans-Williamson Algorithm

A lot of work has been done to accelerate GW algorithms [34]. Based on the "dynamic edge splitting" idea proposed by Cole *et al.* [27] in 2001, Hegde *et al.* [11] recently proposed a fast implementation of the unrooted GW algorithm. We refer to this algorithm as the FGW algorithm. It is currently the fastest heuristic algorithm for PCSTP, and it has a tight approximation guarantee of 2. Here, we use GPrA to improve its solution while maintaining its speed, and the resulting algorithm is referred to as the FGW′ algorithm (Algorithm 5).

There are two phases in the FGW′ algorithm: GW-growth and GW-pruning. In the GW-growth phase, we split each edge $(i,j)$ into two edge parts $ep(i,j)$ and $ep(j,i)$ corresponding to the endpoints $i$ and $j$ (Step 2). The edge splitting method is not clear in Hegde's FGW algorithm in that it has not been specified how the edge is split. In our FGW′ algorithm, we define the edge splitting ratio $s$ ($s \geq 1$) as follows.

$$slack\{ep(i,j)\} = \begin{cases} c(i,j)/s, & i < j \\ (s-1)c(i,j)/s, & i > j \end{cases} \quad (4)$$

The two edge parts $ep(i,j)$ and $ep(j,i)$ share the slack (or cost) of edge $(i,j)$ at the ratio of $1 : (s-1)$. We will later discuss the influence of $s$ on the final solution.

The total number of edge parts is $2|E|$, and the number of edge parts associated with each vertex equals the degree of this vertex. An edge part is active when the vertex it associates with is in an active cluster, otherwise the edge part is inactive. Initially, we set each vertex as a cluster, and the slack of each cluster equals its node weight (compulsory terminals

---

**Algorithm 5** The Proposed FGW′ Algorithm

**Input:** Graph $G(V, E, C, w, c)$, parameter $s$
**Output:** Steiner tree $T \subseteq G$
1: Initialize $T = \emptyset$, the global time $t_g$
2: Split edges into edge parts using Equation (4)
3: Initialize cluster and edge events
4: **while** there are more than one active cluster **do**
5:     Find the closest edge event time $t_e$ and the edge part $ep_1$
6:     Find the closest cluster event time $t_c$ and the cluster $cl$
7:     **if** $t_e \leq t_c$ **then**
8:         Update $t_g$ to $t_e$
9:         Identify the corresponding edge part $ep_2$ to $ep_1$
10:         **if** $ep_1$ and $ep_2$ are in the same cluster **then**
11:             Continue
12:         **else**
13:             Calculate $r$
14:             **if** $r > \mu$ **then** ($\mu$ is a small value close to 0)
15:                 Update the event time of $ep_1$ and $ep_2$
16:             **else**
17:                 Add the corresponding edge to $T$
18:                 Merge the two clusters and their edge parts
19:             **end if**
20:         **end if**
21:     **else**
22:         Update $t_g$ to $t_c$
23:         Deactivate $cl$
24:     **end if**
25: **end while**
26: Remove edges not in the last active cluster from $T$
27: $T = \text{GPrA}(T)$

---

have infinite node weights). All the clusters with a positive slack are active.

There are two types of events, the edge event and the cluster event. The initial slacks of all the active clusters and edge parts are considered as their event time (Step 3). We maintain a global time value $t_g$. As $t_g$ increases, the slacks of active edge parts and clusters decrease. At any time, the remaining slack of an active cluster is the gap between its event time and $t_g$; the remaining slack of an inactive cluster is 0; the remaining slack of an active edge part is the gap between its event time and $t_g$; the remaining slack of an inactive edge part is the gap between its event time and the deactivation time of its cluster.

The edge and cluster events are triggered in the order of their event time. In the cluster event, we simply deactivate the responsible cluster (Step 23). In the edge event, the slack of the responsible edge part is 0 (e.g. $slack\{ep(i,j)\} = 0$). However, the total slack of the responsible edge may not be 0 yet (e.g. $slack\{ep(i,j)\} + slack\{ep(j,i)\} > 0$). Assume that edge part $ep(i,j)$ is the responsible edge part for an edge event. Let $r$ be the slack of edge part $ep(j,i)$. If $r = 0$, then we merge the two clusters connected by edge $(i,j)$ and their edge parts (Steps 17-18; to maintain the speed, even the edge parts between these two clusters are being merged). The slack of the new cluster equals the sum of slacks of the two merged clusters. Suppose the slack of the new cluster is $sl$, we set the event time of the new cluster to be $t_g + sl$. Note that, an inactive cluster may be merged into an active cluster in an edge event. In that case, we need to increase the event time of edge parts in the inactive cluster by the gap between $t_g$ and its deactivation time.

If $r > 0$, then we distinguish two cases to update the event time of these two edge parts (Step 15):

*Case 1:* The cluster containing edge part $ep(j, i)$ is active. Since we expect the slacks of these two edge parts ($ep(i, j)$ and $ep(j, i)$) to become 0 at the same time to trigger a merge event, we split $r$, and update the event time of $ep(i, j)$ and $ep(j, i)$ to $t_g + r/2$.

*Case 2:* The cluster containing edge part $ep(j, i)$ is inactive. We assume that the cluster containing edge part $ep(j, i)$ stays inactive until a merge event is triggered by edge $(i, j)$. Then, for the same reason, we update the event time of $ep(i, j)$ to $t_g + r$, and the event time of $ep(j, i)$ to the deactivation time of its cluster.

Crucially, we update the event time of these two edge parts in the above way so that the two corresponding clusters would be merged in the next event on edge $(i, j)$, assuming both clusters maintain their current activity status. If one of the two clusters changes its activity status, this will not hold. An extreme situation is that both clusters were active and the cluster containing edge part $ep(j, i)$ becomes inactive since. As a result, the next event on edge $(i, j)$ will still have $r > 0$, and we need to split the slack $r$ again. In the worst case, the slack splitting case may keep happening endlessly. However, if we specify a precision value $d$, which means that the slack on an edge can be split for at most $d$ times, then there are at most $O(d)$ events can be triggered on each edge. Notably, the value $d$ is only a theoretical safeguard to maintain the speed. Practically speaking, there are only two events that can be triggered on each edge in most cases. This phenomenon has been observed in both our computational trials and Hegde's work. In our FGW′ algorithm, a small value $\mu$ is used. $\mu$ is close to 0. If $r \leq \mu$ (Step 16), we trigger the merge event. The functions of $d$ and $\mu$ are the same, but $\mu$ is simpler to implement. We end the optimization process above until there is only one active cluster left (Step 4). The subtree in this cluster is the raw solution tree we obtained in the GW-growth phase (Step 26).

In the GW-pruning phase, we prune the raw solution tree above. The Strong Pruning algorithm is used in the GW-pruning phase of the FGW′ algorithm. However, the Strong Pruning algorithm can only find optimal pruning solutions in the rooted instances. Our proposed GPrA is an improvement of the Strong Pruning algorithm. It can find optimal pruning solutions in both rooted and unrooted instances (see Theorem 2). Thus, GPrA is used in the GW-pruning phase of the FGW′ algorithm (Step 27). The FGW′ algorithm has a polynomial time complexity of $O(|E| log |V|)$ and an approximation guarantee of 2 (in the worst case, $O(|E| log |V|)$ is $O(|V|^2 log |V|)$). Note that, it only has a constant approximation guarantee and is not a polynomial time approximation scheme. Therefore, it does not contradict the claim made by Arora *et al.* [29] that a MAX SNP-hard problem does not have a polynomial time approximation scheme, unless P = NP. Its solutions satisfy the following stronger approximation guarantee [11], [35],

$$c(T) + 2w(\overline{T}) \leq 2c(T_{opt}) + 2w(\overline{T_{opt}}) \tag{5}$$

where $T$ is the solution of the FGW′ algorithm, $T_{opt}$ is the optimal solution to PCSTP, $c(T)$ and $c(T_{opt})$ are the total edge costs in $T$ and $T_{opt}$, $w(\overline{T})$ and $w(\overline{T_{opt}})$ are the total node weights not in $T$ or $T_{opt}$. Given a solution of the FGW′ algorithm, a lower bound can be obtained using Equation (5). Since large instances that have not been solved to optimality are used in this paper, it is preferable to use this lower bound to evaluate the solution quality.

### B. The Proposed Fast Implementation of the Rooted Goemans-Williamson Algorithm

The unrooted GW algorithms are faster than their rooted versions. However, it may still be preferable to use the rooted versions in some cases since their solutions are generally better [16]. The FGW′ algorithm proposed above is a fast implementation of the unrooted GW algorithm. In this subsection, we propose its rooted version, which is named as the FGW″ algorithm.

In the FGW″ algorithm, we iteratively select a vertex to be the root. The edge splitting technique is still used in the GW-growth phase, but the cluster containing the root is always inactive. The GW-growth phase terminates when there is no active cluster left. Then, in the GW-pruning phase, GPrA is used to prune the subtree in the root cluster.

In previous work, researchers implement the rooted GW algorithm in instances with compulsory terminals by selecting a compulsory terminal to be the root [16]. However, we observe that it may be preferable to consider all the vertices to be the possible roots, not just compulsory terminals. Consider a triangular graph constructed by vertices $i$, $j$, $k$. Vertices $i$ and $k$ are non-compulsory vertices, and vertex $j$ is a compulsory terminal. Suppose $w(i) = 3, w(j) = 20, w(k) = 20, c(i, j) = 6, c(i, k) = 10, c(j, k) = 11$. Select compulsory terminal $j$ as the root, then the solution is $\{e(i, j), e(i, k)\}$. However, we can obtain the optimal solution $\{e(j, k)\}$ by selecting the non-compulsory vertex $i$ as the root (the raw solution $\{e(i, j), e(j, k)\}$ is obtained in the GW-growth phase, and edge $\{e(i, j)\}$ is pruned in the GW-pruning phase). Thus, even in instances with compulsory terminals, we may prefer to consider every vertex as the possible root. Therefore, the FGW″ algorithm has a time complexity of $O(|E||V| log |V|)$, and an approximation guarantee of $2 - 1/(|V| - 1)$.

The FGW″ algorithm can find better solutions than the FGW′ algorithm in many cases [16]. However, it is currently too slow to apply the FGW″ algorithm in large instances. For example, our computational trials show that the FGW′ algorithm needs around 40 seconds to find a feasible solution in a large instance with 1 million vertices and 10 million edges. Thus, the FGW″ algorithm may need half a million minutes to solve the same instance, which is obviously impractical. Since we focus on designing large and low-cost communication networks, the FGW″ algorithm will not be implemented in this paper. However, it may be preferable to use the FGW″ algorithm in small instances in some cases. Moreover, the optimization process (to try multiple possible roots) of the FGW″ algorithm suits parallel-computing perfectly. Since machines with more and more computing cores are being developed, a parallel FGW″ algorithm could be applied to design large and low-cost communication networks in the future.

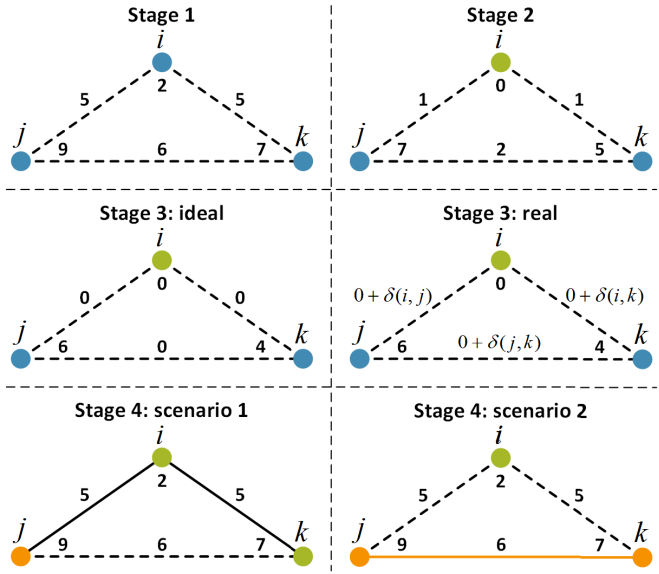

Fig. 4. Applying the Goemans-Williamson approximation scheme to solve a triangular instance composing of three non-compulsory vertices $i$, $j$, $k$. The node weights and edge costs are: $w(i) = 2$, $w(j) = 9$, $w(k) = 7$, $c(i, j) = c(i, k) = 5$, $c(j, k) = 6$. In Stage 1, all the vertices are active (in blue color); in Stage 2, $i$ runs out of slack and becomes inactive (in green color); in the ideal Stage 3, all the three edges run out of slack at the same time; in the real Stage 3, unavoidable computational rounding errors ($\delta$) exist in the slacks of three edges; in the first scenario of Stage 4 where the edge events are processed on $(i, j)$ and $(i, k)$, the solution is $\{j\}$ (in orange color; $(i, j)$ and $(i, k)$ are first included but later pruned); while in the second scenario of Stage 4 where the edge event is processed on $(j, k)$, the solution is $\{(j, k)\}$.

### C. The Solution Certainty of Goemans-Williamson Approximation Scheme

The solution certainty of GW approximation scheme has never been discussed before. Using a particular GW algorithm and a particular implementation method, we always get the same solution for the same instance. In this subsection, we further explain why different solutions can be produced by the same GW algorithm for the same instance when it is implemented differently.

In the GW approximation scheme, cluster or edge events are triggered in the GW-growth phase. The condition where multiple events are triggered simultaneously occurs quite often, and different orders to process these events may induce different solutions. For example, in Figure 4, three edge events are triggered at the same time; if we process the edge events on $(i, j)$ and $(i, k)$ first, then $\{j\}$ is obtained; while if we process the edge event on $(j, k)$ first, then $\{(j, k)\}$ is obtained. Therefore, we have the following theorem.

*Theorem 4: The Goemans-Williamson approximation scheme can produce different solutions for the same*

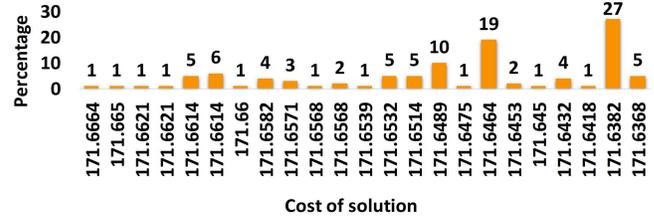

Fig. 5. Different solutions for the handsd01 instance.

instance by processing simultaneously triggered events in different orders.

Our FGW′ algorithm is an improvement of the FGW algorithm. Suppose these two algorithms are implemented on the same instance, and the triggered events are sequenced in the same way, then the outcomes of their GW-growth phase will be the same. However, since GPrA can guarantee the pruning optimality in instances with an arbitrary number of compulsory terminals while the Strong Pruning algorithm cannot, the GW-pruning outcome of our FGW′ algorithm is always better than or equal to that of the FGW algorithm. Therefore, we have the following corollary.

*Corollary 5: When the triggered events are sequenced in the same way, the solution of our FGW′ algorithm always dominates that of the FGW algorithm.*

In the fast implementation of GW approximation scheme, the cluster and edge events are stored in priority queues. Thus, it is impossible to manipulate the sequence of triggered events since the time complexity would be ruined. In this paper, we give our own code for the FGW′ algorithm, and the triggered events are sequenced differently from those in the FGW algorithm in Hegde's paper [11]. As a result, although our FGW′ algorithm almost always produces better solutions, the solutions of the FGW algorithm in Hegde's paper are better than those of our FGW′ algorithm in a few benchmark instances. However, we consider this as acceptable since the solutions of our FGW′ algorithm are still better than theirs in the large majority of benchmark instances. Moreover, in this paper, we also give our own code for the FGW algorithm, and the triggered events are sequenced in the same way with those in our FGW′ algorithm. Consequently, the solutions of our FGW′ algorithm dominate those of the FGW algorithm in all the benchmark instances, which verifies the corollary above.

Furthermore, it must be mentioned that the computational rounding errors in practice may induce different solutions by changing the sequence of simultaneously triggered events. For example, Equation (4) is used to split edges in the FGW′ algorithm, and small computational rounding errors may be produced in this process. Let us consider the triangular instance in Figure 4. In the real Stage 3, suppose the total slacks on the edges are respectively $0 + \delta(i, j)$, $0 + \delta(i, k)$, $0 + \delta(j, k)$, where $\delta$ is the error of the slack on each edge. If $\delta(i, j) = \min\{\delta(i, j), \delta(i, k), \delta(j, k)\}$, then we will process the event on edge $(i, j)$ first. Different values of $s$ in Equation (4) may induce different computational rounding errors, and then induce different solutions. For example, in Figure 5, we implement the FGW′ algorithm with random values of $s$ multiple times, and dozens of different solutions are obtained. Clearly, the computational rounding errors induced by different values of $s$ are unavoidable in practice. Thus, we have the following conclusion.

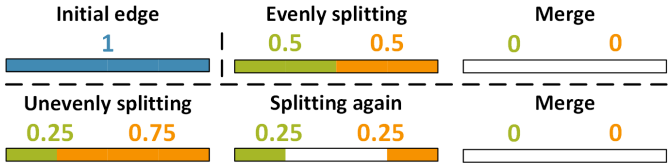

Fig. 6. $s = 2$ induces the least number of triggered events.

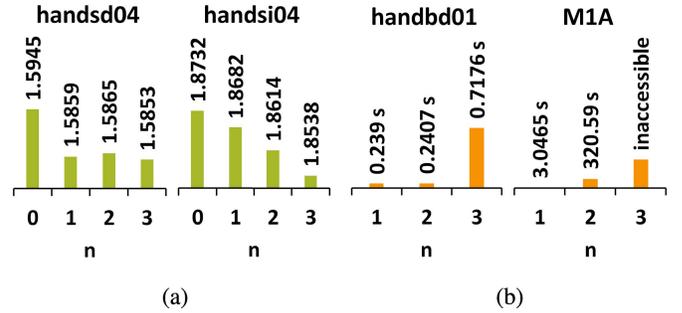

Fig. 7. The trade-off of $n$ in the post-process. (a) The approximation ratio. (b) The running time.

*Conclusion 1: The FGW′ algorithm with different values of $s$ may produce different solutions for the same instance.*

Notably, similar issues widely exist for many other algorithms in practice. For example, Dijkstra's algorithm [36] is widely used to find the shortest path in a graph. There may be multiple shortest paths, and the one it finds depends on the specific codes. Consequently, the algorithms built on Dijkstra's algorithm may produce different solutions for the same instance. On the other hand, all the solutions produced by the FGW′ algorithm satisfy the strong approximation guarantee of Equation (5). It is impractical to predict that which value of $s$ induces the best solution since it requires the knowledge of the induced computational rounding errors and a detailed analysis of the specific instance. However, we observe that the different solutions induced by different values of $s$ have similar qualities in practice: the net-cost of the worst solution is at most 0.43% larger than that of the best one when implementing the FGW′ algorithm with 100 random values of $s$ for each of the well-known Hand instances (the 95% Confidence Intervals of this percentage are [0.09%, 0.14%]).

Moreover, we observe that different values of $s$ can also change the number of triggered events in the FGW′ algorithm. For example, in Figure 6, suppose the edge is connected to two active clusters; if $s = 2$, which means that we evenly split the edge initially, then the slacks of these two edge parts will run out at the same time, and the first event on this edge will directly induce the merge of two adjacent clusters; however, if we unevenly split the edge initially, then the slacks of these two edge parts will run out at different times, and a second event on this edge is required before the merge of two adjacent clusters. Therefore, $s = 2$ induces the least number of triggered events in the FGW′ algorithm. This has also been verified in our computational trials. In this paper, we try 100 random values of $s$ for each of the well-known Hand instance, and then select the best one to obtain dominating solutions to those in Hegde's paper (note that, the edge splitting method is not clear in Hegde's paper), while for the newly generated larger instances, since it is impractical to predict the solutions induced by different values of $s$ and also too slow to implement a large number of different values of $s$, we set $s = 2$ to minimize the number of triggered events.

## VII. EXPERIMENTS

It is challenging to design large and low-cost communication networks. Our heuristic algorithms and post-processing techniques can help meet this challenge. In this section, we indicate the competitiveness of our FGW′ and MSTG algorithms by comparing them with the state-of-the-art algorithms in the largest existing benchmark instances and some newly generated instances that are even larger. Moreover, we apply our P3 to update the best-known solution for a notoriously difficult benchmark instance to show that it can improve near-optimal solutions to PCSTP. Remarkably, even though large instances with millions of vertices are used, all the computational trials are conducted on a commonly used personal computer from 2016 (Intel Core i7-4790 CPU with 3.60GHz), and the consumed memory in the largest instance is less than 8 GB. The reported running times are averaged over 10 trials for each instance. The time spent on inputting and outputting data is excluded from the reported running time, but the time spent on all the other processes is included. The codes and our later generated M instances are available at [37].

### A. The Trade-Off of $n$ in the Tree Growing Algorithm (TGA)

The time complexity of TGA is $O(|V|^{n+1})$, where $n$ is the upper bound of the lengths of possible path candidates that are checked. Clearly, TGA with a larger $n$ has a higher probability of improving a suboptimal solution, but at the cost of a longer running time. Here, we investigate this trade-off through computational trials.

First, we implement P3 with different values of $n$ to post-process suboptimal solutions of the MSTG algorithm in two benchmark instances: handsd04 and handsi04. The improvements are shown in Figure 7a, where the approximation ratios are ratios of the post-processed solutions to the optimal ones [23]. It can be seen from handsi04 that a larger $n$ induces a larger improvement on the suboptimal solution. Nevertheless, this is not always true. For example, in handsd04, the post-processed solution when $n = 2$ has a larger approximation ratio than that when $n = 1$. The reason is that, when $n$ is large, a long branch may be added to take up the opportunity of adding a better short branch. Since it is computationally too expensive to compare the candidate branches before adding them, the scenario above may be unavoidable. Nonetheless, it is still preferable to set $n$ large to give TGA a greater chance to improve suboptimal solutions.

On the other hand, a large $n$ makes TGA slow. We implement P3 with different values of $n$ to post-process suboptimal solutions of the MSTG algorithm in two benchmark instances: handbd01 ($|V| = 169,800$, $|E| = 338,551$) and M1A ($|V| = 1,000,000$, $|E| = 10,000,000$). Clearly, the running time of P3 grows quickly as $n$ increases. Therefore, even though a larger $n$ has a greater chance to improve suboptimal solutions, it is not recommended to set $n$ too large to make P3 slow. In the following computational trials, we set $n = 2$ in the Hand instances and $n = 1$ in the M instances to make such a trade-off between speed and solution quality.

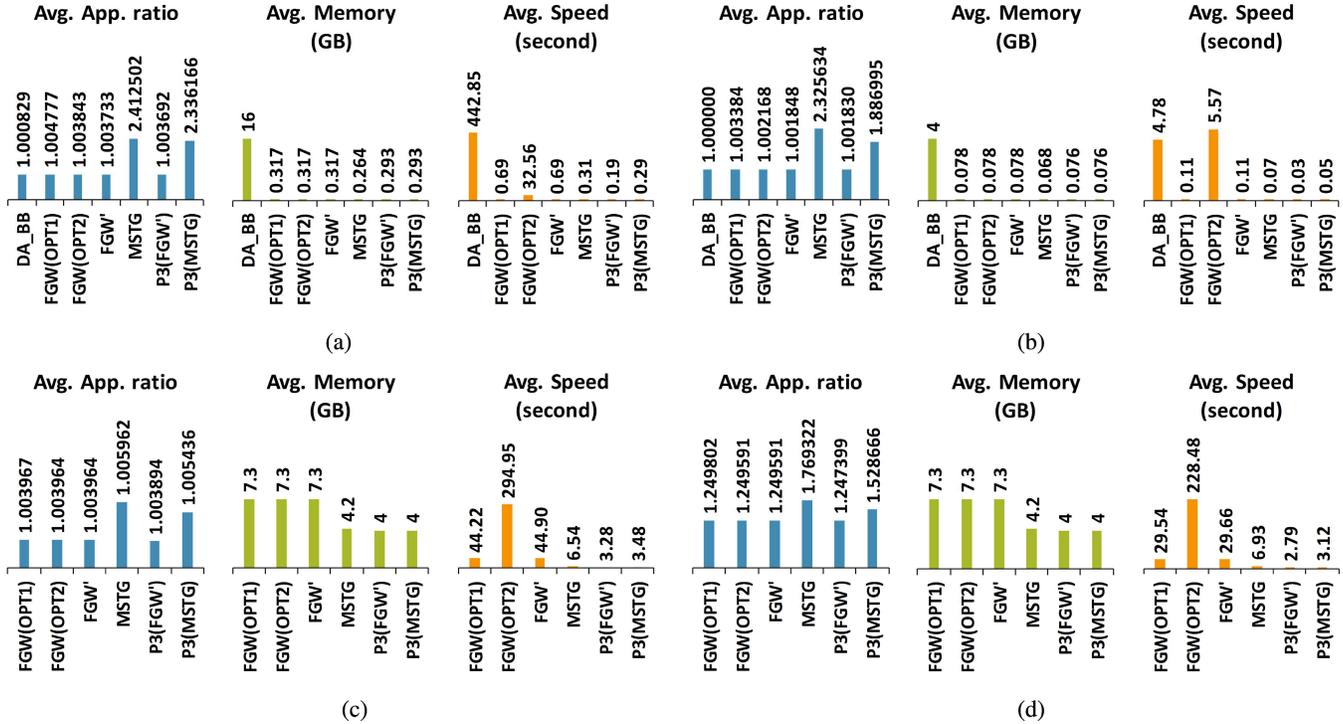

Fig. 8. The computational trials in the existing Hand instances and the newly generated M instances. (a) handb instances ($|V| \approx 169,800$, $|E| \approx 338,551$). (b) hands instances ($|V| \approx 42,500$, $|E| \approx 84,475$). (c) M_A instances ($|V| = 1,000,000$, $|E| = 10,000,000$). (d) M_B instances ($|V| = 1,000,000$, $|E| = 10,000,000$).

### B. Application to Benchmark Instances

Our FGW' and MSTG algorithm can find fast high-quality solutions to PCSTP in large instances. Here, we indicate their competitiveness by comparing them with the state-of-the-art algorithms for both the largest existing benchmark instances and some newly generated instances that are even larger. We also use P3 to post-process their solutions.

The Hand instances generated by Hegde *et al.* [11] in 2014 are currently the largest benchmark instances for PCSTP, and they are available at [22]. The largest Hand instance has 169,800 vertices and 338,551 edges. Some communication networks may be much larger than this. For example, the telecommunication network in a large city like Melbourne may need to cover millions of buildings, and a local 5G wireless network may connect millions of mobile devices. Thus, we generate some new larger instances, the M instances, to reflect such a large scale. There are 1 million vertices and 10 million edges in each M instance. In each of M1A to M20A, all the vertices have positive node weights, while in each of M1B to M20B, only 10,000 vertices have positive node weights, and the node weights of the other vertices are 0. We apply both the Hand and M instances here.

Since Hegde's FGW algorithm [11] is the state-of-the-art heuristic algorithm for PCSTP. We compare it with our FGW' and MSTG algorithm in the same settings. Johnson's Strong Pruning algorithm [16] is used in the pruning phase of Hegde's FGW algorithm. In instances with no compulsory terminal, like our applied instances and most other ones for PCSTP [11], [14], roots need to be randomly selected in the Strong Pruning algorithm to obtain good pruning results [16]. We provide two options here: randomly select 1 root to prune once; and randomly select 100 roots to prune 100 times. The resulting FGW algorithms are referred to as FGW(OPT1) and FGW(OPT2). Moreover, DA_BB [23] is the state-of-the-art exact algorithm for PCSTP. Thus, we also compare it with our FGW' and MSTG algorithm in the Hand instances. However, since the state-of-the-art exact algorithms are not our focus; they cannot solve large instances that our algorithms are designed to solve; and most computers, including ours, are not powerful enough to implement them due to the small RAM size, we do not implement DA_BB in this paper, but only compare its computational results in [23], where a much more powerful computer is used.

The detailed computational results are shown in the supplement, while their statistic evaluations are shown in Figure 8, where three evaluation standards are used: 1) the average approximation ratio to the optimal solution or the best-known lower bound to date [23]; 2) the average memory consumption; 3) the average speed. Clearly, DA_BB has the highest solution quality, but it consumes a lot more memories and are much slower than the other heuristic algorithms (its running times are reported in [23], where a much more powerful computer is used; its 16 GB memory consumption in the handb instances is also reported in [23], while its 4 GB memory consumption in the hands instances is estimated based on the CPLEX memory estimation guideline [38]). The other state-of-the-art exact algorithms, such as Staynerd [20], have similar demands on computational resources with DA_BB. Since the RAM sizes of most machines in the market are currently below 16 GB, it is reasonable to say that the Hand instances are the largest instances that the state-of-the-art exact algorithms can solve at present. Hence, they may not be able to design larger communication networks. For example, it can be estimated from Figure 8 that DA_BB may consume hundreds of GBs memory to solve the M instances, which is impractical in most cases.

On the other hand, FGW(OPT1) has a similar speed with our FGW′ algorithm, while FGW(OPT2) is much slower than the two algorithms above. The reason is that only one root root is selected in FGW(OPT1) and our FGW′ algorithm can directly find the optimal root, but FGW(OPT2) need to select 100 random roots. Consequently, the raw solution tree we obtained in the GW-growth phase only need to be pruned once in FGW(OPT1) and our FGW′ algorithm, but it needs to be pruned 100 times in FGW(OPT2). Even so, since there may be hundreds of thousands of root candidates and it is thus impractical to select all of them, our FGW′ algorithm has a dominating solution quality over both FGW(OPT1) and FGW(OPT2). Furthermore, it may be worth mentioning that, since the GW approximation scheme can obtain different solutions for the same instance (see Section VI-C), the solutions of the FGW algorithm in Hegde's paper [11] are better than those of FGW(OPT1) and FGW(OPT2) in some instances. Nevertheless, by trying more root candidates, FGW(OPT1) and FGW(OPT2) can obtain the same solutions of our FGW′ algorithm, which dominate those of the FGW algorithm in Hegde's paper. In respect of our MSTG algorithm, even though there is no approximation guarantee for it, its approximation ratios are small in these instances, which means that it can find reasonably high-quality solutions in practice. Moreover, it is quasilinear and thus is much faster than the other GW algorithms in the M instances, and its consumed memory is also smaller.

P3 has improved the FGW′ and MSTG solutions in the large majority of these instances, and its running time is small and neglect-able. We have also shown the improvements on approximation ratios in Figure 9. Note that, since the speed of FGW(OPT1) is at the same level with FGW′ + P3, we show the improvements from FGW(OPT1) to our FGW′ and P3(FGW′), but not from FGW(OPT2), which is much slower. Since the solutions of these GW algorithms are very close to the optimal ones, the improvements on them are small by percentage. Nevertheless, the costs for these improvements, i.e., the memory consumption and running time of FGW′ and P3, are also small. Moreover, as large communication networks may be expensive in practice, a small percentage of improvement may still mean that a high cost is saved. For example, in large telecommunication networks that worth billions of dollars (e.g. [39]), a 0.1% improvement can save millions of dollars. Furthermore, in larger instances where only the MSTG algorithm is fast enough to be implemented, P3 can improve its solutions significantly.

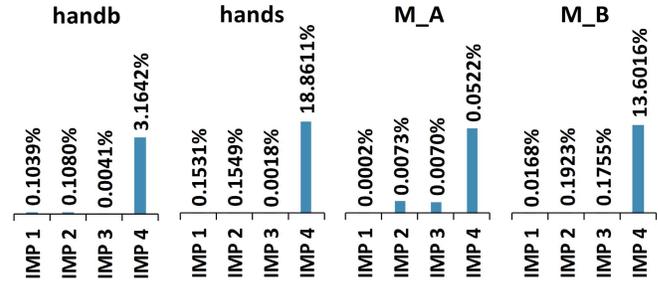

Fig. 9. The improvements on approximation ratios. IMP 1 is the improvement from FGW(OPT1) to FGW′; IMP2 is that from FGW(OPT1) to P3(FGW′); IMP3 is that from FGW′ to P3(FGW′); and IMP4 is that from MSTG to P3(MSTG).

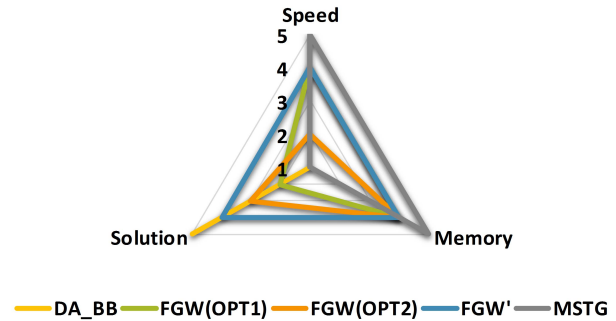

Fig. 10. Rating of the state-of-the-art algorithms.

Ultimately, we rate these algorithms in Figure 10 to clearly show the advantages of our FGW′ and MSTG algorithms over the state-of-the-art algorithms in designing large and low-cost communication networks. Furthermore, we summarize the following conclusions from these computational trials:

- the state-of-the-art exact algorithms may not be able to design large and low-cost communication networks due to their low speed and high demand on computational resources.

- our FGW′ algorithm produces better solutions than the state-of-the-art heuristic algorithm: FGW, while keeping a high speed and a low demand on computational resources (FGW′ keeps the time complexity of $O(|E| log |V|)$, while FGW must increase it to $O(|E| log |V| + |V|^2)$, by trying all the pruning roots, to find these better solutions).

- our MSTG algorithm has a higher speed and a lower demand on computational resources than the other algorithms, and its solutions have small approximation ratios in practice. Therefore, it has advantages over the other algorithms in time or computational resource sensitive scenarios, and it can solve extremely large instances that the other algorithms are too slow to solve.

- our P3 can improve near-optimal solutions to PCSTP (the best-known solution for a Hand instance is updated later by P3).

TABLE I
OUR IMPROVED SOLUTIONS OF STAYNERD AND DA_BB

| Solver | Instance | Solu. | P3_Solu. | P3_Time |
|---|---|---|---|---|
| Staynerd | handbd04 | 1820.073 | 1819.8188 | 0.2737s |
| Staynerd | handbd06 | 1533.5899 | 1533.4144 | 0.2793s |
| Staynerd | handbi02 | 532.6165 | 532.4787 | 0.2773s |
| Staynerd | handbi04 | 3226.9192 | 3226.8756 | 0.2618s |
| Staynerd | handbi06 | 2930.6424 | 2930.5869 | 0.2642s |
| Staynerd | handbi13 | 4.2682 | 4.2676 | 0.2385s |
| Staynerd | handsd02 | 160.3458 | 160.3386 | 0.0634s |
| Staynerd | handsd04 | 494.554 | 494.5349 | 0.0619s |
| DA_BB | handbi13 | 4.274497 | 4.274367 | 0.2426s |

*C. Post-Process Suboptimal Solutions of the State-of-the-Art Exact Algorithms*

The state-of-the-art exact algorithms may only be able to find suboptimal solutions in large instances. Staynerd [20] and DA_BB [23] are the state-of-the-art exact algorithms for PCSTP. Here, we show that P3 can also improve their suboptimal solutions in the Hand instances. The numbers of solutions of Staynerd and DA_BB that have not been proven to be optimal in the Hand instances are respectively 21 and 3. We implement P3 to improve these solutions (they are provided by Ivana Ljubić and Martin Luipersbeck, who proposed Staynerd and DA_BB with others). The improved solutions are shown in Table I. It can been seen that P3 has improved 8 suboptimal solutions of Staynerd and 1 suboptimal solution of DA_BB. The running time of P3 in these instances is around 0.25s. Therefore, it is reasonable to conclude that P3 can improve not only suboptimal solutions of our heuristic algorithms, but also those of the state-of-the-art exact algorithms. Hence, it may be preferable to combine P3 with the state-of-the-art exact algorithms together to solve PCSTP in some cases. Note that, the Staynerd solution of handbi13 we received from Ivana Ljubić and Martin Luipersbeck is better than that in their previously published paper [20]. However, P3 can still improve it. The improved solution (4.2676) is the new best-known solution for this instance.

## VIII. CONCLUSION

It is challenging to design large and low-cost communication networks. In this paper, we formulate this challenge as the Prize-Collecting Steiner Tree Problem (PCSTP). We first propose some effective post-processing techniques to improve suboptimal solutions to PCSTP. Then, based on these techniques, we propose two heuristic algorithms for PCSTP. We indicate the competitiveness of our heuristic algorithms over the state-of-the-art ones in both the largest existing benchmark instances and some newly generated instances that are even larger. Moreover, we apply our post-processing techniques to update the best-known solution for a notoriously difficult benchmark instance to show that they can improve near-optimal solutions to PCSTP. In summary, we demonstrate that our heuristic algorithms and post-processing techniques can help design large and low-cost communication networks.

## ACKNOWLEDGMENT

The authors would like to express their sincere thanks to I. Ljubić and M. Luipersbeck for providing the solutions of Staynerd and DA_BB for the Hand instances. They also sincerely thank the anonymous reviewers for providing valuable comments to improve this work.

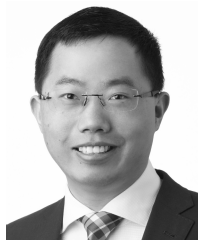

**Yahui Sun** received the master's and bachelor's degrees in aerospace engineering from the Harbin Institute of Technology in 2012 and 2014, respectively, and the Ph.D. degree from the University of Melbourne in 2018. He is currently a Post-Doctoral Fellow with the Research School of Engineering, Australian National University. His research interests are computer networks and network data analysis, especially Steiner tree problems in node-weighted graphs and their applications to social, biomedical, telecommunication, and wireless sensor networks.

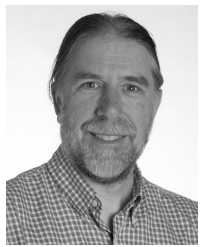

**Marcus Brazil** is currently an Associate Professor and also a Reader with the Department of Electrical and Electronic Engineering, The University of Melbourne. His main research interest is in optimal network design, with applications to telecommunications, wireless sensor networks, VLSI physical design, underground mining, and infrastructure for electric vehicles. He also combines this with more theoretical work, particularly in the area of Steiner Trees. His research and consultancy work in optimization for underground mine design has been particularly successful, having received strong support from a number of Industry bodies, such as BHP Billiton and Newmont Australia Ltd.

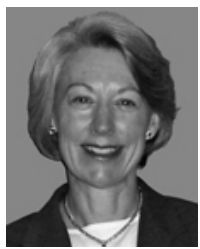

**Doreen Thomas** was the Head of the School of Electrical, Mechanical and Infrastructure Engineering, University of Melbourne. She is currently a Professor Emeritus with the Department of Mechanical Engineering, The University of Melbourne. She has applied her fundamental mathematical research in network optimization to applications in a number of areas, including underground mine access. The software encapsulating her work has been licensed to mining companies, such as BHP Billiton, Newmont, and Rio Tonto to reduce underground mine development and haulage costs. She is a Fellow of the Australian Academy of Technology and Engineering, and also a Fellow of Engineers Australia.

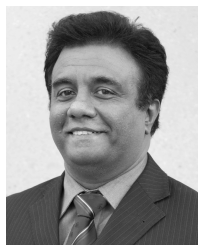

**Saman Halgamuge** (F'17) was the Director of the Research School of Engineering, Australian National University. He is currently a Professor with the Department of Mechanical Engineering, University of Melbourne. His research interests are in machine learning, including deep learning, big data analytics, and optimization. These applications vary from smart grids and sustainable energy generation to bioinformatics and neuro-engineering. His fundamental research contributions are in Big Data Analytics with unsupervised and near unsupervised type learning as well as in Transparent Deep Learning and Bioinspired Optimization. He was a member of the Australian Research Council College of Experts for Engineering, Information and Computing Sciences.